\documentclass[twocolumn]{webofc}
\usepackage[varg]{txfonts}   
\usepackage{lineno}
\pdfpageattr {/Group << /S /Transparency /I true /CS /DeviceRGB>>}
\newcommand{\cwglong}{Auger\,--\,Telescope Array\xspace}
\newcommand{\cwg}{Auger\,--\,TA\xspace}
\newcommand{\xmax}{\ensuremath{X_\mathrm{max}}\xspace}
\newcommand{\gcm}{\ensuremath{\mathrm{g\,cm^{-2}}}\xspace}
\newcommand{\qgsii}{QGSJetII-04\xspace}
\newcommand{\eposlhc}{EPOS-LHC\xspace}
\newcommand{\sibyll}[1]{Sibyll~#1\xspace}
\newcommand{\energyr}[2]{\ensuremath{10^{#1}~{\rm eV} < E < 10^{#2}~{\rm eV}}\xspace}
\newcommand{\arange}[2]{\ensuremath{#1^\circ-#2^\circ}}
\newcommand{\InTA}{$\otimes$~TA\xspace}
\newcommand{\meanXmax}[1]{\ensuremath{\langle
    X^{#1}_{\mathrm{max}}\rangle}}
\newcommand{\sigmaXmax}[1]{\ensuremath{\sigma(X^{#1}_{\mathrm{max}})}}
\begin{document}
\title{Depth of maximum of air-shower profiles: testing the
  compatibility of measurements performed at the Pierre Auger
  Observatory and the Telescope Array experiment}

\author{\firstname{Alexey} \lastname{Yushkov}\inst{1}
  \and
  \firstname{Jose} \lastname{Bellido}\inst{2}
  \and
  \firstname{John} \lastname{Belz}\inst{3}
  \and
  \firstname{Vitor} \lastname{de Souza}\inst{4}
  \and
  \firstname{William} \lastname{Hanlon}\inst{3}
  \and
  \firstname{Daisuke} \lastname{Ikeda}\inst{5}
  \and
  \firstname{Pierre} \lastname{Sokolsky}\inst{3}
  \and
  \firstname{Yoshiki} \lastname{Tsunesada}\inst{6}
  \and
  \firstname{Michael} \lastname{Unger}\inst{7}
  \lastname{~for the Pierre Auger}\inst{8}\fnsep\thanks{\email{auger_spokespersons@fnal.gov}}
  \lastname{and Telescope Array}\inst{9}\fnsep\thanks{\email{whanlon@cosmic.utah.edu}} collaborations
}

\institute{
  Institute of Physics of the Czech Academy of Sciences, Prague, Czech Republic
  \and
  University of Adelaide, Adelaide, S.A., Australia
  \and
  High Energy Astrophysics Institute \& Department of Physics and Astronomy, University
  of Utah, Salt Lake City, USA
  \and
  Instituto de F\'isica de S\~ao Carlos, Universidade de S\~ao Paulo, S\~ao Carlos, Brasil
  \and
  Institute for Cosmic Ray Research, University of Tokyo, Kashiwa, Chiba, Japan
  \and
  Graduate School of Science, Osaka City University, Osaka, Osaka, Japan
  \and
  Karlsruhe Institute of Technology, Institut f\"ur Kernphysik, Karlsruhe, Germany
  \and
  Observatorio Pierre Auger, Av. San Mart\'in Norte 304, 5613, Malarg\"ue, Argentina\\
  Full author list: \url{http://www.auger.org/archive/authors_2018_10.html}
  \and
  Telescope Array Project, 201 James Fletcher Bldg, 115 S. 1400 East, Salt Lake City, UT $84112-0830$, USA\\
  Full author list: \url{http://www.telescopearray.org/research/collaborators}
}

\abstract{%

  At the Pierre Auger Observatory and the Telescope Array, the
  measurements of depths of maximum of air-shower profiles, \xmax, are
  performed using direct observations of the longitudinal development
  of showers with the help of the fluorescence telescopes. Though the
  same detection technique is used at both installations, the
  straightforward comparison of the characteristics of the measured
  \xmax distributions is not possible due to the different approaches
  to the analysis of the recorded events. In this work, the \cwglong
  composition working group presents a technique to compare the \xmax
  measurements from the Auger Observatory and the Telescope
  Array. Applying this technique the compatibility of the first two
  moments of the measured \xmax distributions is qualitatively tested
  for energies $10^{18.2}~{\rm eV} < E < 10^{19.0}~{\rm eV}$ using the
  recently published Telescope Array data from the Black Rock Mesa and
  Long Ridge fluorescence detector stations. For a quantitative
  comparison, simulations of air showers with \eposlhc, folded with
  effects of the Telescope Array detector, are required along with the
  inclusion in the analysis of the systematic uncertainties in the
  measurements of \xmax and the energies of the events.}
\maketitle
\section{Introduction}

The large amount of data on the ultra-high-energy cosmic rays,
collected by the Pierre Auger Observatory~\cite{augerfull_NIMA2015}
operating in the Southern Hemisphere since 2004 and the Telescope
Array (TA)~\cite{TA_ISVHECRI2006} operating in the Northern Hemisphere
since 2008, opens unprecedented possibilities for performing a
comparison of the spectra, arrival directions and mass compositions of
the cosmic rays coming from the complementary regions of the sky. Any
differences in the characteristics of the primary radiation found in
such type of analysis would have important astrophysical implications
and would help to clarify the origin of the most energetic particles
in the Universe. For an unequivocal attribution of such differences to
factors having an astrophysical nature, the comparison of the data
of Auger and TA should rely on a deep understanding of the systematic
effects present in the measurements of the two experiments which
employ different kinds of the surface detectors (SD), different
atmospheric monitoring equipment and programs, different designs of
the fluorescence detectors (FD) and eventually different approaches to
the data analysis. To address these issues a close collaboration of
the people from various experiments is required, and to this scope,
several working groups were created in preparation for the
ultra-high-energy cosmic ray symposium that took place in Geneva in
2012~\cite{uhecr2012}.

The activities of the mass composition working group in the recent
years have been focused on the comparison of the Auger and TA data on
the depth of the maximum of the extensive air showers measured with
the use of the fluorescence telescopes. Until the present time, a good
agreement between the two data sets was found as regarding the
evolution of \meanXmax{} with energy~\cite{cwg_uhecr2016}, so
regarding the compatibility of the shapes of the \xmax
distributions~\cite{cwg_icrc2017_xmax} (the latter analysis was done
for energies below 10~EeV). In 2018 the TA Collaboration has published
new data~\cite{TA_ApJ2018_xmax} including results on the first two
moments of the \xmax distributions with the larger statistics than
ever before. In this paper we present a preliminary qualitative
analysis of the compatibility of these TA measurements with the data
of the Pierre Auger Observatory and outline the next steps required
for finalizing the comparison.

\section{Data}
\label{sec:data}

\subsection*{Pierre Auger Observatory}

At the Pierre Auger Observatory the longitudinal development of air
showers is measured with the FD consisting of 24 fluorescence
telescopes covering each $30^\circ$ in azimuth and \arange{1.5}{30} in
elevation. The telescopes are grouped in units of six at four sites
around the SD array of the area of 3000~km${}^2$. There is an
additional site with three high elevation telescopes for detection of
air showers with energies $E\lesssim10^{18}$~eV, but events with such
energies will not be discussed in this paper.

An extensive program of monitoring of atmospheric parameters,
important for a proper reconstruction of \xmax and energy of air
showers, is run at the Observatory (see~\cite{augerfull_NIMA2015} for
more details). Pressure, humidity and temperature at different
atmospheric depths are determined on a 3-hour basis using data from
the Global Data Assimilation System. Vertical aerosol optical depth
(VAOD) is measured hourly with the FD using light profiles of the
laser shots from two central laser facilities. The presence of clouds
in the field of view of the FD telescopes is controlled each
15~minutes with the help of cloud cameras and elastic lidars,
additional information on cloud coverage is obtained using laser shots
from the laser facilities and data from the Geostationary Operational
Environmental Satellites. For events of the extremely high energies or
having unusual longitudinal profiles, a rapid monitoring of
atmospheric conditions is performed using the lidars and the
Photometric Robotic Atmospheric Monitor.

\begin{figure}[h]
  \centering
  \includegraphics[width=0.42\textwidth]{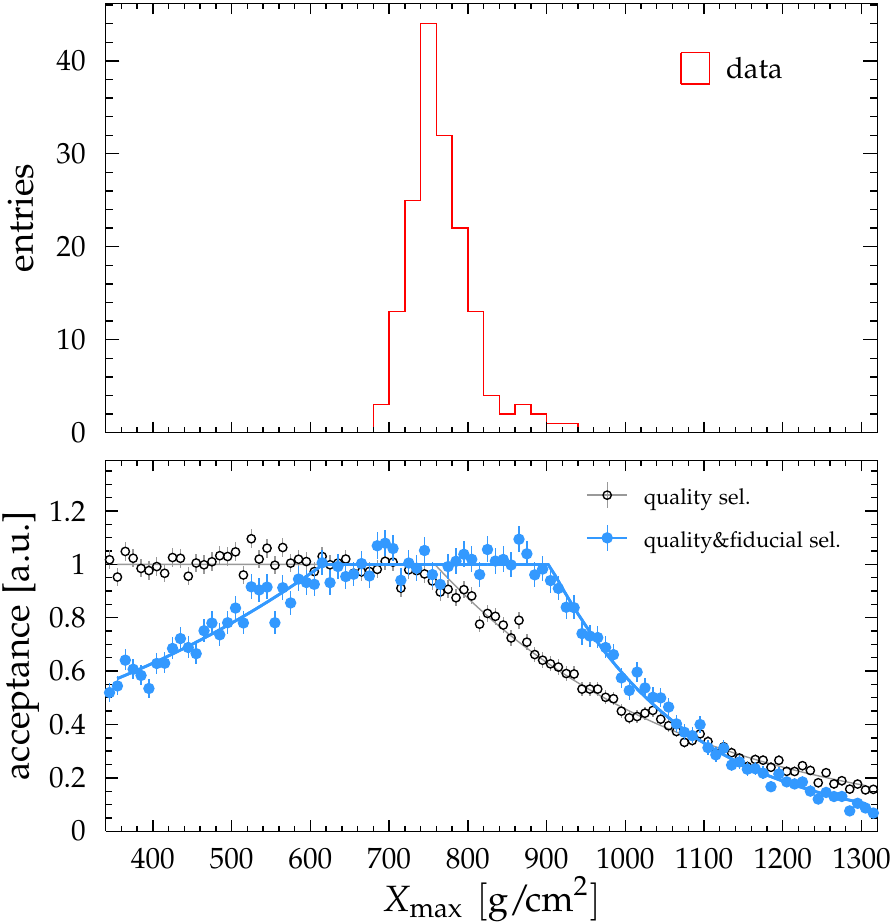}
  \caption{Bottom: relative acceptance of the Auger FD for
    \energyr{19.0}{19.1} with (blue solid circles) and without (black
    open circles) fiducial field-of-view
    selection~\cite{longXmax2014}. With the fiducial field-of-view
    selection almost the whole measured \xmax distribution (top panel)
    is in the constant acceptance region.}
  \label{fig:accept}
\end{figure}

The reconstruction of the recorded events is performed using the Auger
Offline software framework~\cite{offline07}. Only hybrid events,
i.e. the events having at least one triggered SD station, that satisfy
a number of quality selection criteria, are included in the \xmax
analysis. Further, the high quality data pass an additional fiducial
field-of-view selection to guarantee an unbiased acceptance of the
showers almost independently on their \xmax (see
Fig.~\ref{fig:accept}). Finally, the data are corrected for the
reconstruction, residual acceptance biases and resolution effects, and
thus the resulting \meanXmax{} and \sigmaXmax{} can be compared
directly to the predictions from the ideal (without any detector) MC
simulations (see~\cite{longXmax2014} for full details). The most
recent Auger data~\cite{auger_icrc2017_xmax} come from the period
$12/2004-12/2015$ and contain about 10550 events in the energy range
$E>10^{18.2}$~eV discussed further.

\begin{figure}[h]
  \centering
  \includegraphics[width=0.48\textwidth]{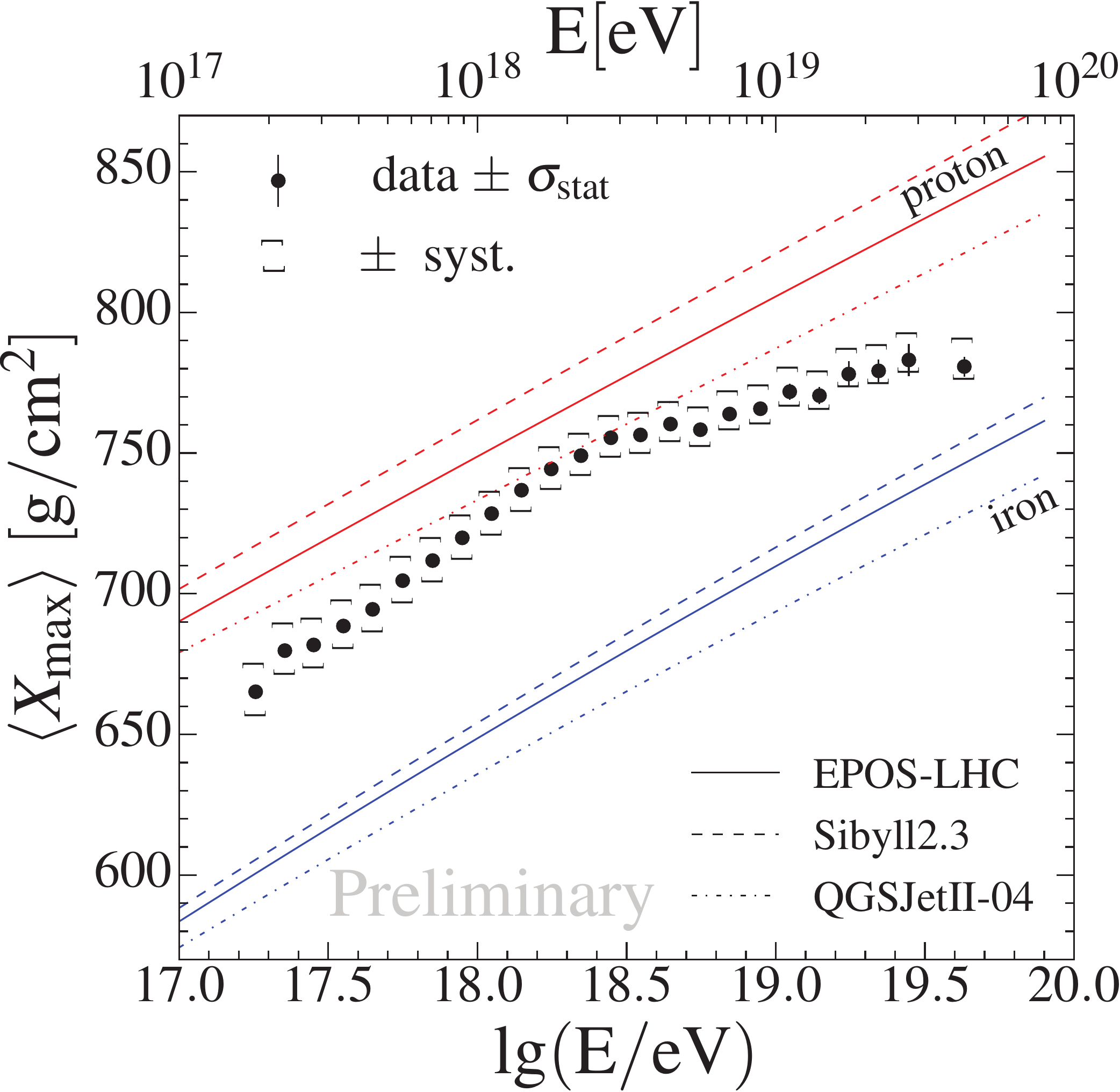}\\[0.2cm]
  \includegraphics[width=0.48\textwidth]{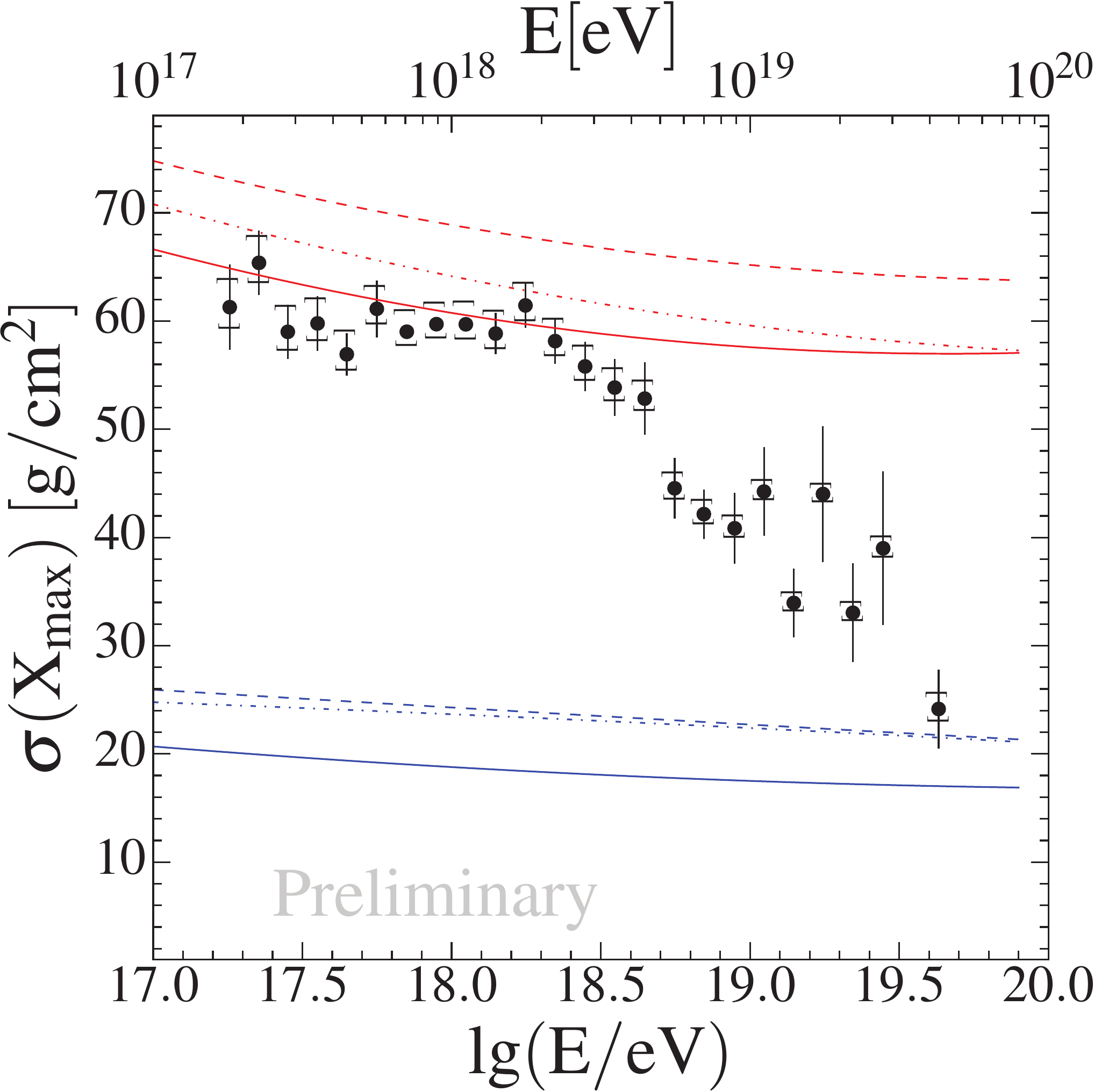}
  \caption{Measurements of \meanXmax{} (top) and \sigmaXmax{} (bottom)
    at the Pierre Auger Observatory~\cite{auger_icrc2017_xmax}
    compared to the predictions for proton and iron nuclei of the
    hadronic models \eposlhc, \sibyll{2.3} and \qgsii.}
  \label{fig:auger_moments}    
\end{figure}

In Fig.~\ref{fig:auger_moments} one can see the evolution of the first
two \xmax moments measured with the Auger Observatory. Starting from
lower energies up to about 2~EeV, \meanXmax{} in the data is
approaching the MC predictions for protons, thus the primary mass
composition is becoming lighter. For energies above 2~EeV, the behavior
of both \meanXmax{} and \sigmaXmax{} is consistent with the increase
of the average mass in the primary beam~\cite{unger_mass_uhecr2018}.

\subsection*{Telescope Array experiment}

The TA data discussed in this paper were collected by the
fluorescence telescopes installed at the Black Rock Mesa (BR) and Long
Ridge (LR) sites~\cite{TA_ApJ2018_xmax}. There are 12 fluorescence
telescopes at each of the sites covering the total field-of-view of
$108^\circ$ in azimuth and \arange{3}{33} in elevation. These FD
telescopes are placed in the southern part of the TA SD
of area of 700~km${}^2$.

The properties of the atmospheric profiles at the TA site are obtained
from the data of the Global Data Assimilation System and, for
estimating the systematic uncertainties in the measurements of \xmax,
from the data collected with the NOAA National Weather Service
radiosondes. The VAOD is measured each 30 minutes using laser shots
from the central laser facility. The reconstruction of \xmax is
performed with an average $\rm{VAOD} = 0.04$ for all nights and the
VAOD measurements are then used for estimation of the systematic
uncertainties of \meanXmax{} and \sigmaXmax{}. The cloud coverage and
cloud thickness are judged by eye and logged hourly by operators in
the field.

The TA events used in the \xmax analysis are hybrid, as in the
case of Auger. For an event to be accepted it is required the
triggering of both the FD and of three SD counters adjacent to each
other. The details on the reconstruction and selection of the events
can be found elsewhere~\cite{TA_ApJ2018_xmax}. Differently from the
Auger analysis, no fiducial field-of-view selection or corrections for
the reconstruction and acceptance biases are applied to the TA
data. Consequently, a larger number of events is kept in the data
set, but the comparison of the data to the predictions of the
interaction models can be done only using the MC simulations processed
through the same analysis chain as the data and including thus the
analysis biases and the effects of the TA detector.  The period of the
data taking is $05/2008-11/2016$, the data set consists of 3330 events
with $E>10^{18.2}$~eV.

\begin{figure}[h]
  \centering
  \includegraphics[width=0.48\textwidth]{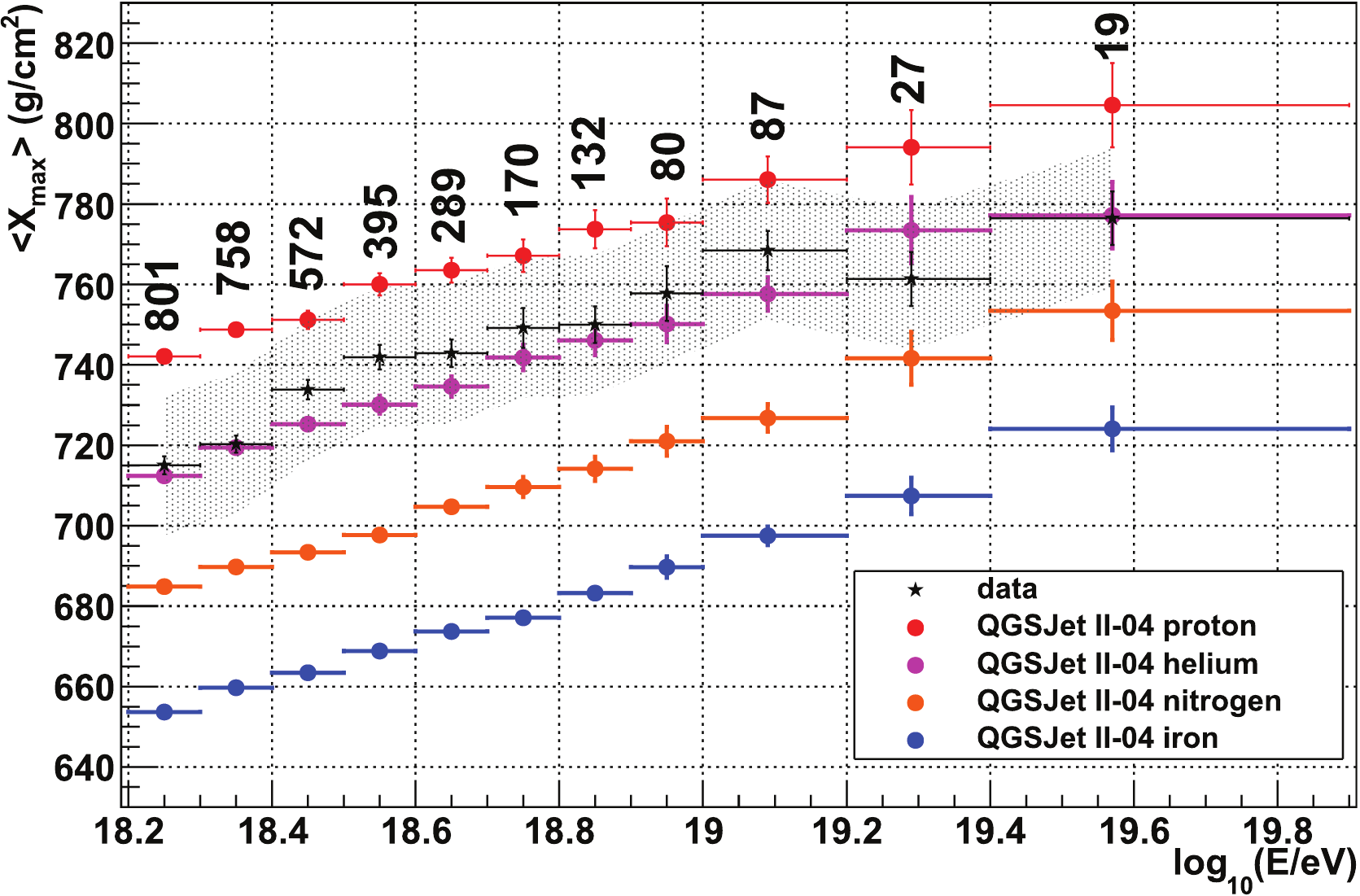}
  \caption{Measurements of \meanXmax{} at the TA
    BR/LR~\cite{TA_ApJ2018_xmax} compared to the predictions for
    pure beams of the hadronic model \qgsii.}
  \label{fig:ta_moments}    
\end{figure}

From the comparison of \meanXmax{} measured by the TA BR/LR to the
predictions of the \qgsii model (Fig.~\ref{fig:ta_moments}) and from
the analysis of the shapes of \xmax
distributions~\cite{TA_ApJ2018_xmax, hanlon_mass_uhecr2018}, it follows
that the data are compatible within the statistical and systematic
errors to pure protons for all energies $E>10^{18.2}$~eV. Due to the
lower detector acceptance for deep high energy showers and due to the
low number of the detected events, for $E>10^{19.0}$~eV the shapes of
the measured \xmax distributions become compatible also to the MC
predictions for other pure primary compositions (helium,
nitrogen)~\cite{TA_ApJ2018_xmax, hanlon_mass_uhecr2018}.

\subsection*{Remarks on the Auger and TA $\boldsymbol\xmax$ measurements}
\begin{figure}[h]
  \centering
  \includegraphics[width=0.44\textwidth]{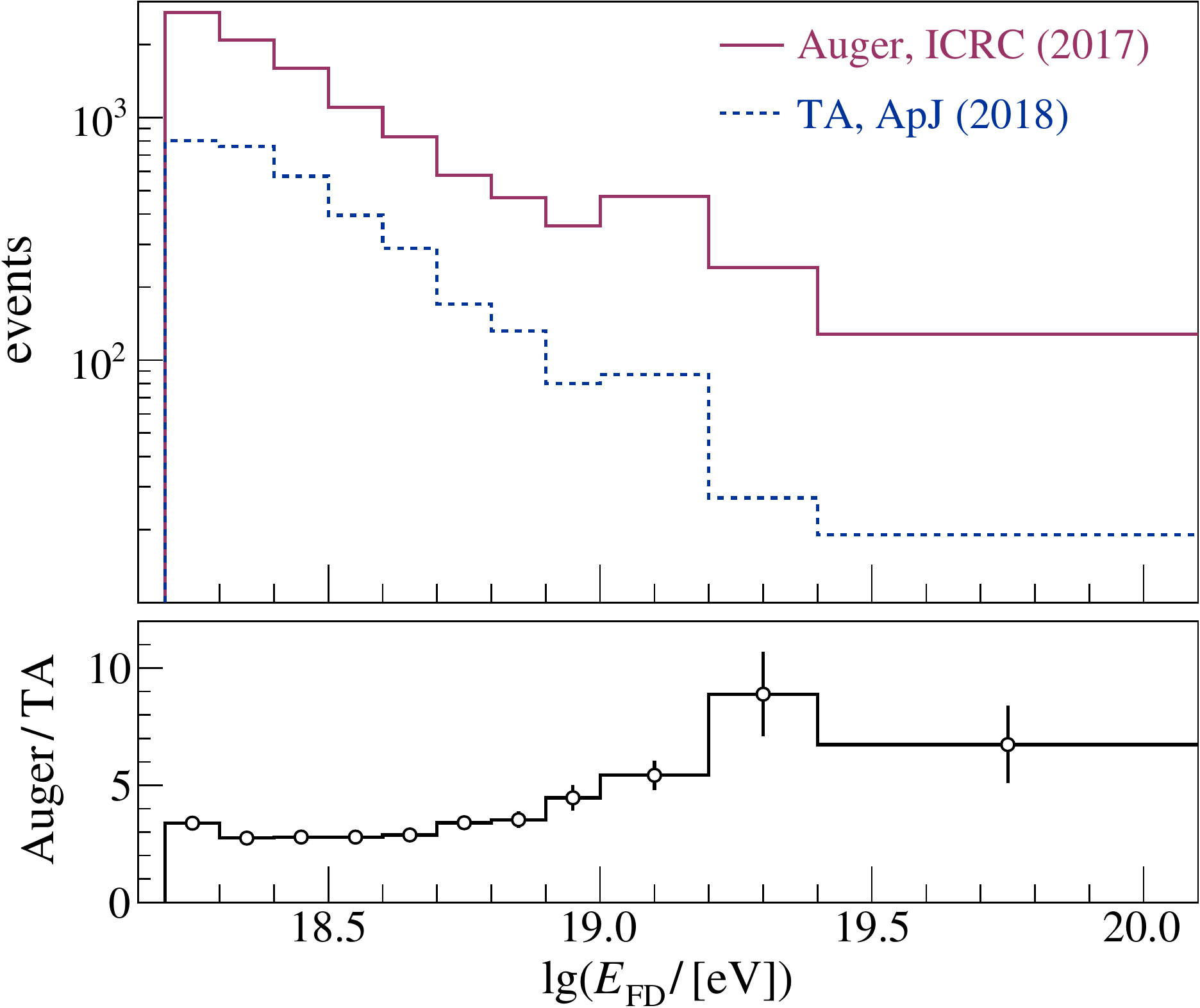}
  \caption{Numbers of events (top) and their ratio (bottom) in the
    \xmax data sets of Auger~\cite{auger_icrc2017_xmax} and TA
    BR/LR~\cite{TA_ApJ2018_xmax} for the common energy range
    $E>10^{18.2}$~eV. The energy binning is the same as used by the
    TA~\cite{TA_ApJ2018_xmax}. Data are binned using the energy
    scales of each of the experiments.}
  \label{fig:Nevts}    
\end{figure}

For energies above $\sim10^{18.5}$~eV the first two moments, the shape
of \xmax distributions and the elongation rate of \meanXmax{} measured
by Auger are incompatible to the MC predictions for
protons~\cite{longXmax2014,auger_icrc2017_xmax} (the same conclusions have
been obtained from the measurements involving the information from the
SD~\cite{auger_mix_2016, augerDelta2017}). The interpretation stating
that the TA \xmax measurements are compatible to the predictions for
\qgsii protons~\cite{TA_ApJ2018_xmax, hanlon_mass_uhecr2018} does not
necessarily contradict the Auger results. In case of the TA BR/LR
data, only the compatibility to the pure beams was tested, thus it can
not be excluded that the TA data can be described well by the mixed
compositions of \qgsii (as was already shown
in~\cite{cwg_icrc2017_xmax}) and of other interaction models. It
should be noted as well that the statistical significance of such a
comparison will be lower than in case of Auger due to the difference
in the sizes of the data sets (Fig.~\ref{fig:Nevts}) reflecting the
difference in hybrid exposures (sizes of the SD arrays, data taking
periods and detection/selection efficiencies) of the Auger and TA
experiments.

Instead of comparing the interpretations, the primary aim of the
composition working group is the comparison of the \xmax measurements
themselves, including the check of the compatibility of the first two
\xmax moments, of the shapes of the \xmax distributions and of the
\meanXmax{} elongation rates. Further we present the technique for
doing such a comparison, perform a preliminary qualitative
comparison of \meanXmax{} and \sigmaXmax{} and outline the next steps
needed for a completion of the full quantitative evaluation of the
compatibility of the Auger and TA \xmax measurements.

\section{A method of comparison of the Auger and TA $\boldsymbol\xmax$ measurements}

As explained in the previous section, the Auger and TA \xmax
measurements cannot be compared directly because of the use of
different approaches to the analysis of the recorded events. In the case
of Auger, due to the application of the fiducial field-of-view
selection and consequent correction of the biases, the \xmax moments
are free from the detector effects and can be compared directly to the
ideal MC simulations. The data of the TA on the other hand include all
detector and analysis biases.

To compare the Auger and TA data one should convert Auger \xmax values
into the values folded with the TA detector effects (Auger
\xmax\InTA). To do this, it was proposed~\cite{cwg_uhecr2012} to find
the MC compositions giving the best description of the shapes of the
Auger \xmax distributions and to process these compositions through
the full simulations and analysis chain used for the TA BR/LR
data. The MC compositions describing the Auger \xmax distributions
were found in~\cite{longXmax_fits2014,auger_icrc2017_xmax} and will be
referred hereafter as AugerMix.

\begin{figure}[h]
  \centering
  \includegraphics[width=0.35\textwidth]{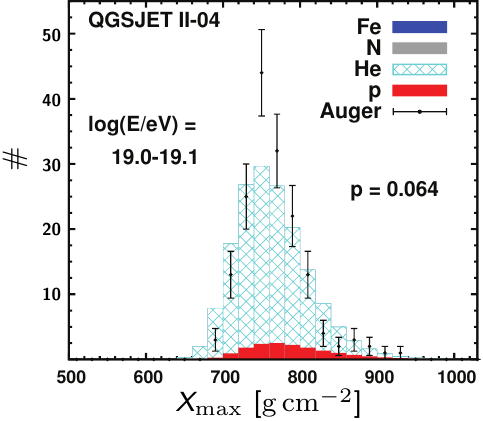}\\[0.2cm]
  \includegraphics[width=0.35\textwidth]{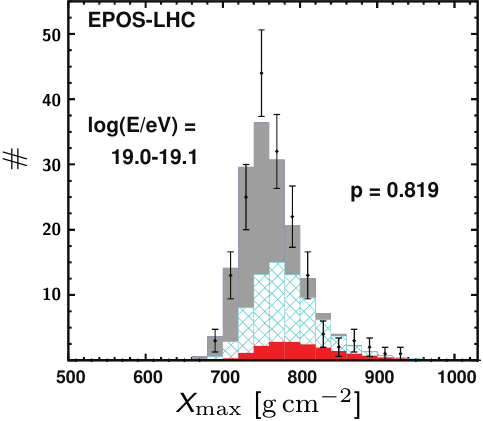}
  \caption{AugerMix for \qgsii (top) and \eposlhc (bottom) for
    \energyr{19.0}{19.1}~\cite{longXmax2014}. The agreement with the
    data is indicated using $p\/$-values.}
  \label{fig:fracfit}
\end{figure}

In Fig.~\ref{fig:fracfit} an example of AugerMix compositions for
\qgsii and \eposlhc in comparison to the Auger data is shown for the
energy bin \energyr{19.0}{19.1}. One can see that the MC template for
\eposlhc provides a better description of the data compared to the
template for \qgsii. As demonstrated in~\cite{auger_icrc2017_xmax},
the $p\/$-values characterizing the agreement between AugerMix of \qgsii
and the Auger data are $\sim0.01$ in all energy bins between 1 and
10~EeV. The reason for a non-satisfactory description of the data
might be that the AugerMix of \qgsii is composed only of protons and
helium and the widths of the \xmax distributions for such a mix of
light elements are too wide compared to the widths found in
data~\cite{auger_icrc2017_xmax}\footnote{c.f. to the conclusions
  of~\cite{auger_mix_2016} that independently on the interaction model
  the presence of elements heavier than helium is required to describe
  data on the correlation between \xmax and signal in the SD
  stations.}.

\begin{figure*}[h]
  \centering
  \includegraphics[width=0.4\textwidth]{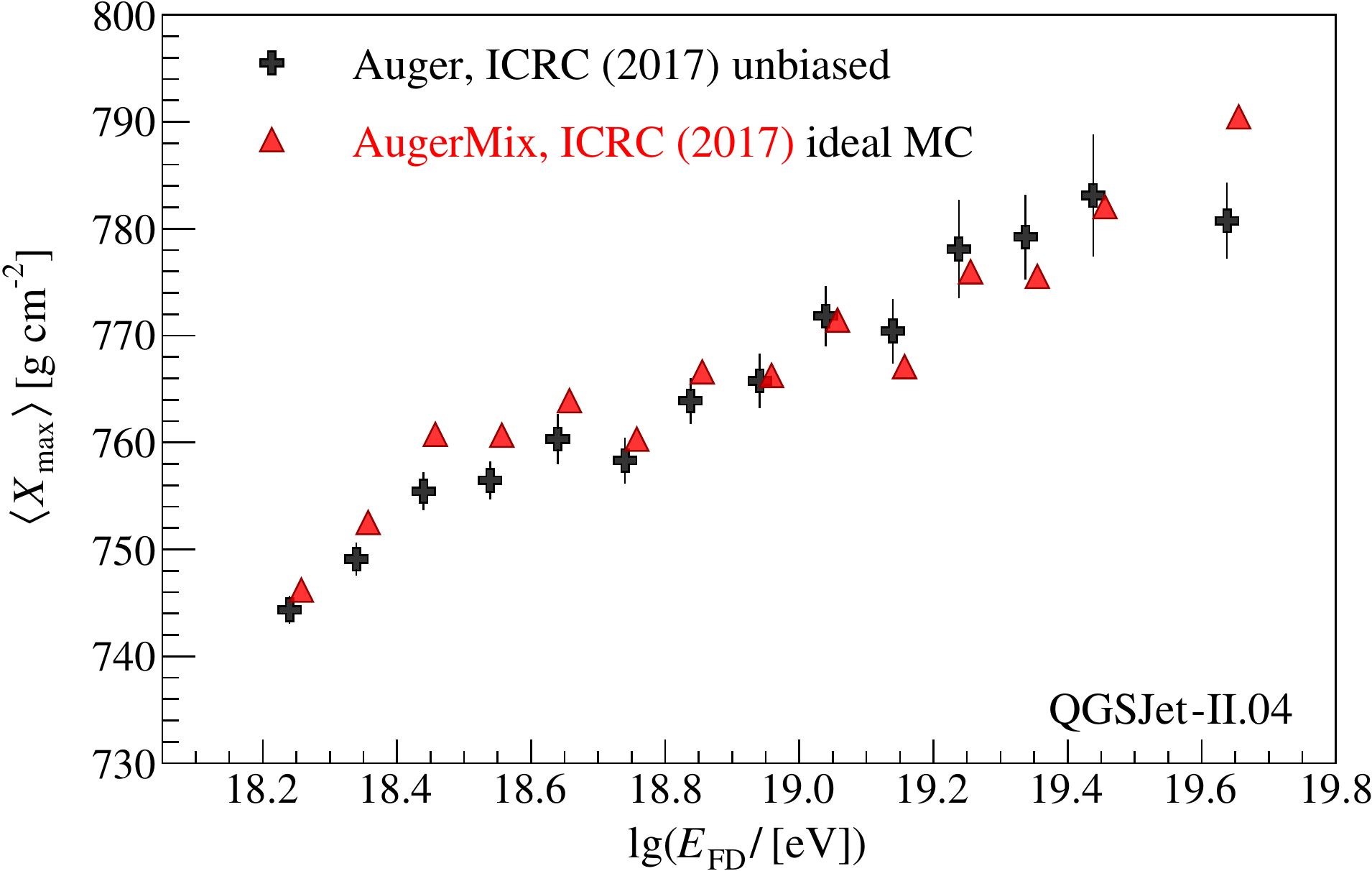}
  \includegraphics[width=0.4\textwidth]{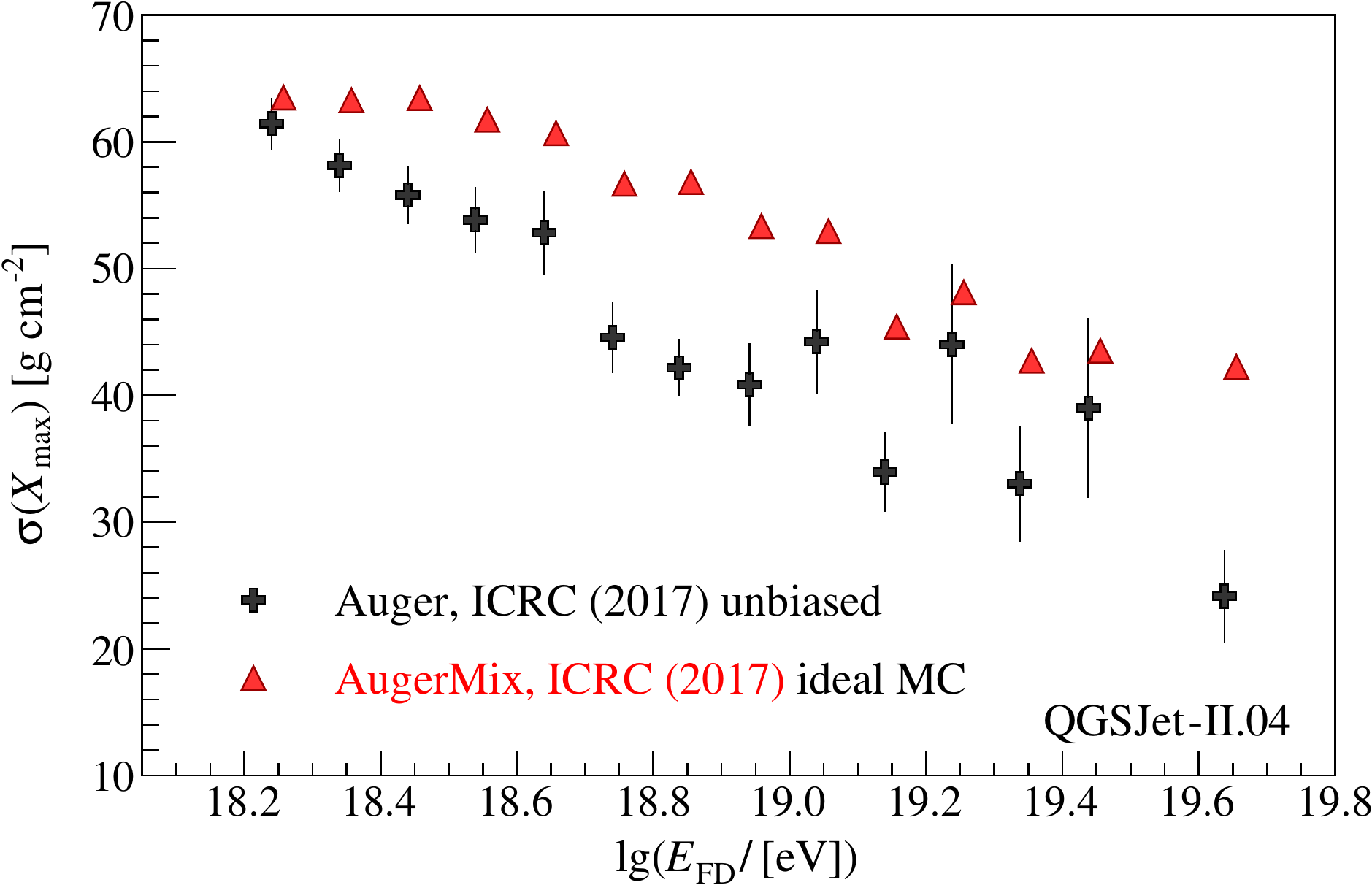}\\[0.2cm]
  \includegraphics[width=0.4\textwidth]{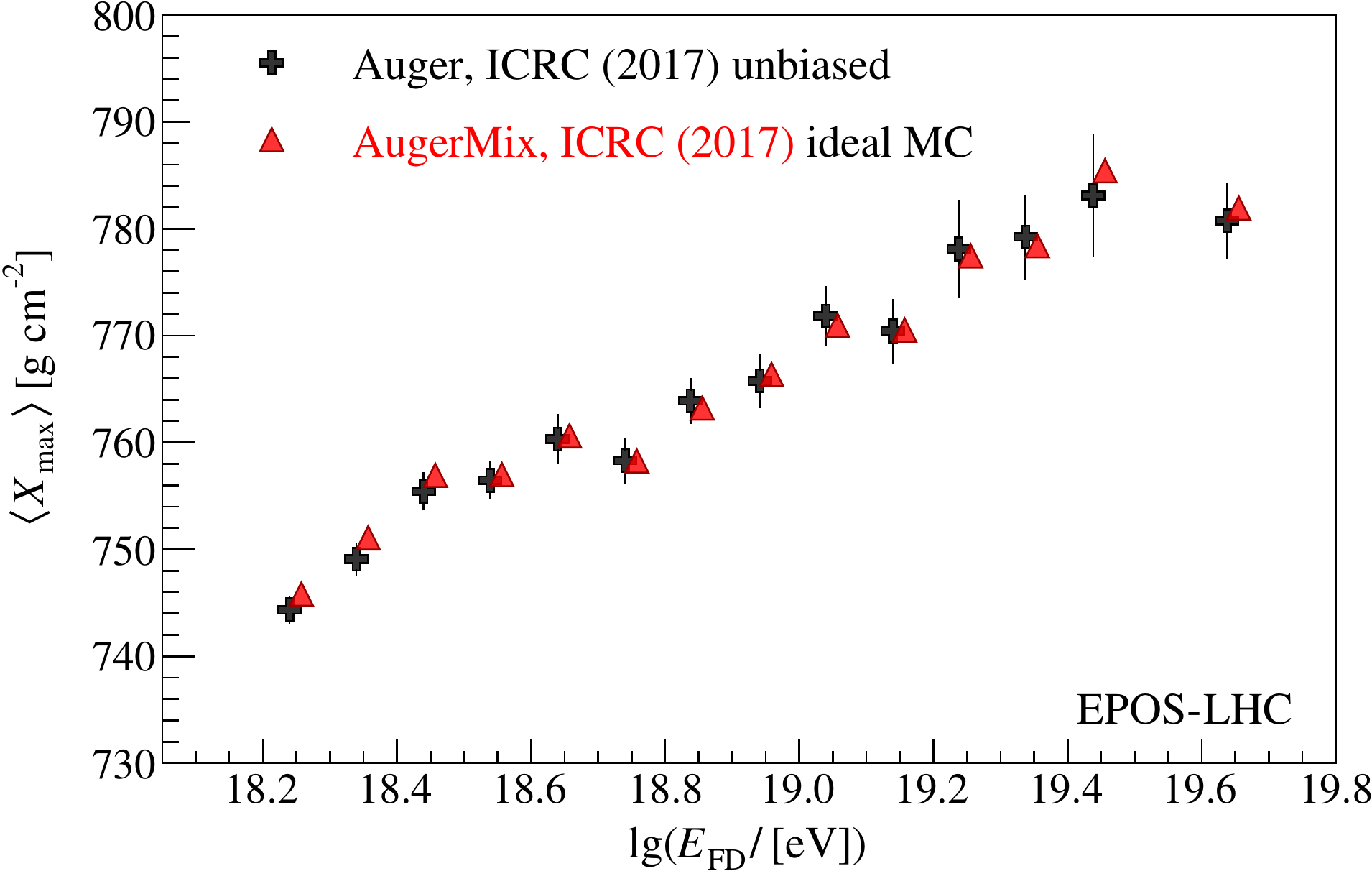}
  \includegraphics[width=0.4\textwidth]{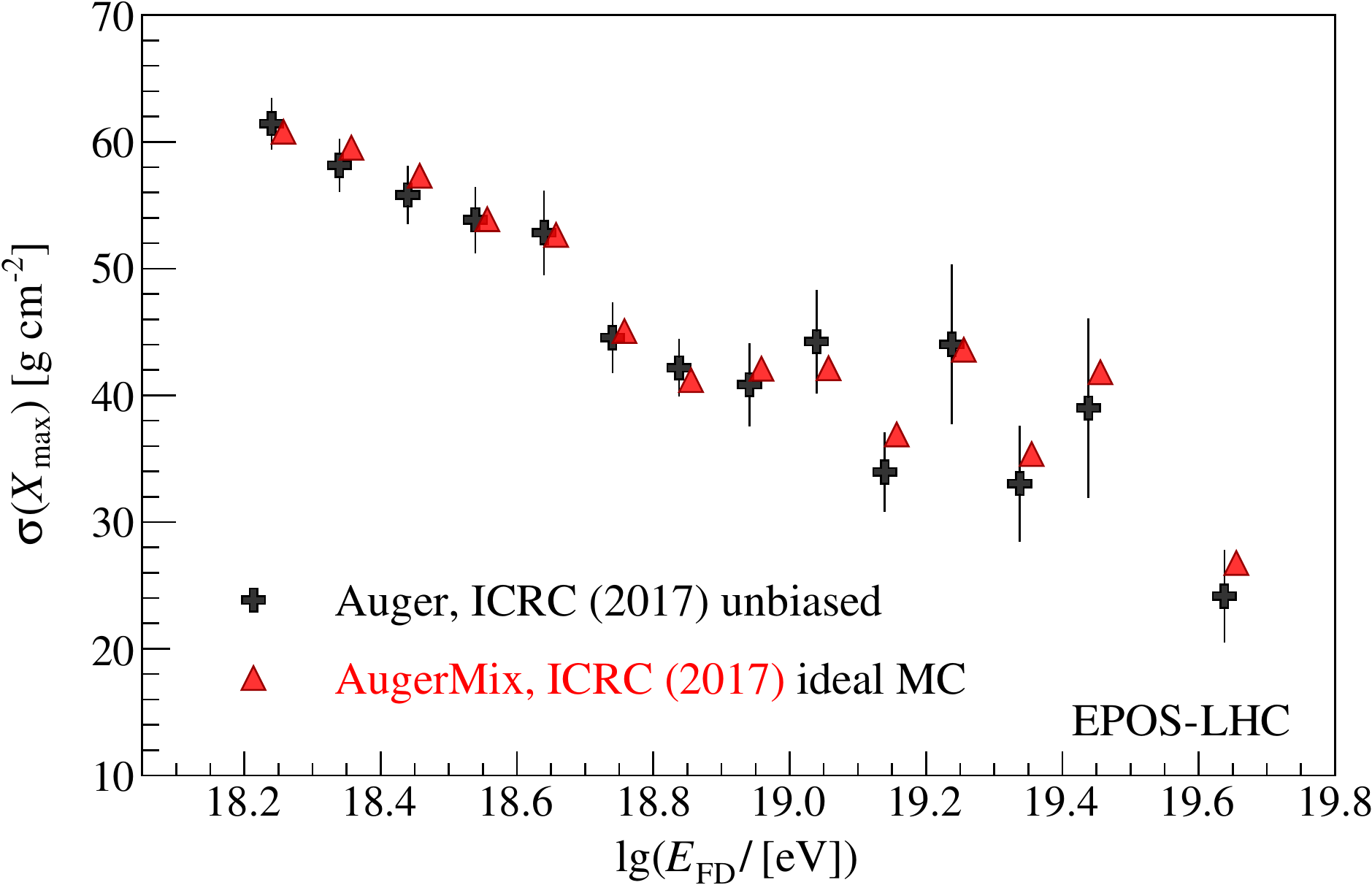}
  \caption{Comparison of \meanXmax{} and \sigmaXmax{} in the Auger
    data and in the AugerMix compositions obtained with \qgsii (top)
    and \eposlhc (bottom)~\cite{auger_icrc2017_xmax}.}
  \label{fig:mcdata_moments}
\end{figure*}

This fact is illustrated in Fig.~\ref{fig:mcdata_moments} where one
can see that \sigmaXmax{} for AugerMix of \qgsii is significantly
larger compared to the values found in the data. Since the AugerMix
compositions of \qgsii do not reproduce correctly the shapes of the
\xmax distributions measured by Auger, they are not the optimal
choice for performing the comparison of the Auger and TA
measurements. For the proper comparison, one can use the mass
composition templates found using \eposlhc: as shown in
Fig.~\ref{fig:mcdata_moments}, with this interaction model both
measured \meanXmax{} and \sigmaXmax{} are described well, as also
indicated by the good $p\/$-values found in~\cite{auger_icrc2017_xmax}
testing the agreement between the \xmax distributions.

At the moment of writing of this paper the only available simulations
for the TA BR/LR detector were done with \qgsii while \eposlhc
simulations have not been finished yet\footnote{these simulations are
  to be finished before the end of 2018.}. Thus for a preliminary
comparison of the \xmax moments, one can only resort to a re-weighting
procedure first introduced and thoroughly tested
in~\cite{cwg_uhecr2016}. The re-weighting allows one to transform
\xmax distributions for the AugerMix of \qgsii into the \xmax
distributions of the AugerMix of \eposlhc taking all reconstruction,
selection biases and TA detector effects into account. Technically
this is achieved by the re-weighting of each reconstructed event of
\qgsii by a factor equal to the ratio of the \xmax p.d.f.'s
AugerMix(\eposlhc)/AugerMix(\qgsii) estimated at the true (generated)
\xmax and energy of this event.

\begin{figure}[h]
  \centering
  \includegraphics[width=0.42\textwidth]{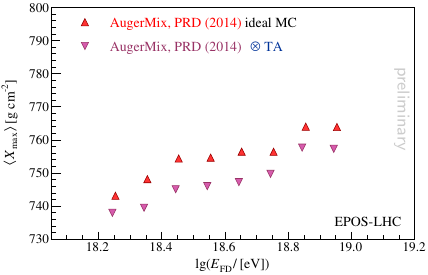}\\[0.2cm]
  \includegraphics[width=0.42\textwidth]{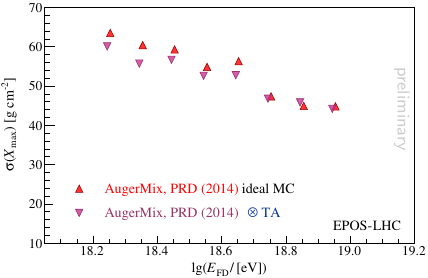}
  \caption{\meanXmax{} (top) and \sigmaXmax{} (bottom) for the
    AugerMix composition of \eposlhc at the generation level (ideal
    MC) and after the simulations and analysis for the
    TA BR/LR detector (\InTA).}
  \label{fig:moments_inta_mc}
\end{figure}

For the comparison presented here the re-weighting was performed using
the AugerMix compositions from~\cite{longXmax2014} for the energy
range \energyr{18.2}{19.0}. From Fig.~\ref{fig:moments_inta_mc} one can
see that \meanXmax{}~\InTA are $\approx5~\gcm$ smaller than the
generated values. This bias can be caused by the relatively smaller
detector acceptance for the showers in the deeper tails of the \xmax
distributions. Similarly, \sigmaXmax{}~\InTA is also a few \gcm smaller
than in the ideal MC.

\begin{figure*}[h]
  \centering
  \includegraphics[width=0.47\textwidth]{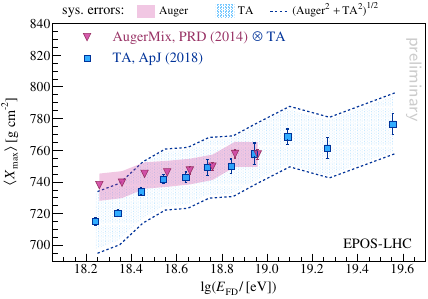}
  \includegraphics[width=0.47\textwidth]{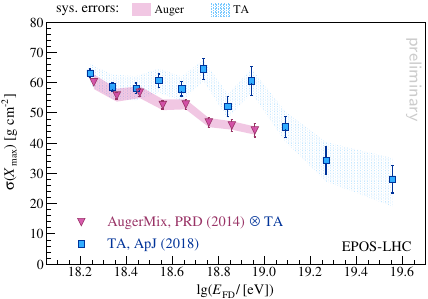}
  \caption{Comparison of \meanXmax{} (left) and \sigmaXmax{} (right)
    of AugerMix~\InTA for \eposlhc (representing the Auger data
    transferred into the TA detector) to the TA BR/LR 
    measurements~\cite{TA_ApJ2018_xmax}. Error bars show the
    statistical errors, shaded areas~---~the systematic uncertainties
    of each of the experiments, dashed lines in the \meanXmax{}
    plot~---~systematic uncertainties of the Auger and TA \xmax
    measurements combined in quadrature.}
  \label{fig:data_moments_inta}
\end{figure*}

\begin{figure}[h]
  \centering
  \includegraphics[width=0.45\textwidth]{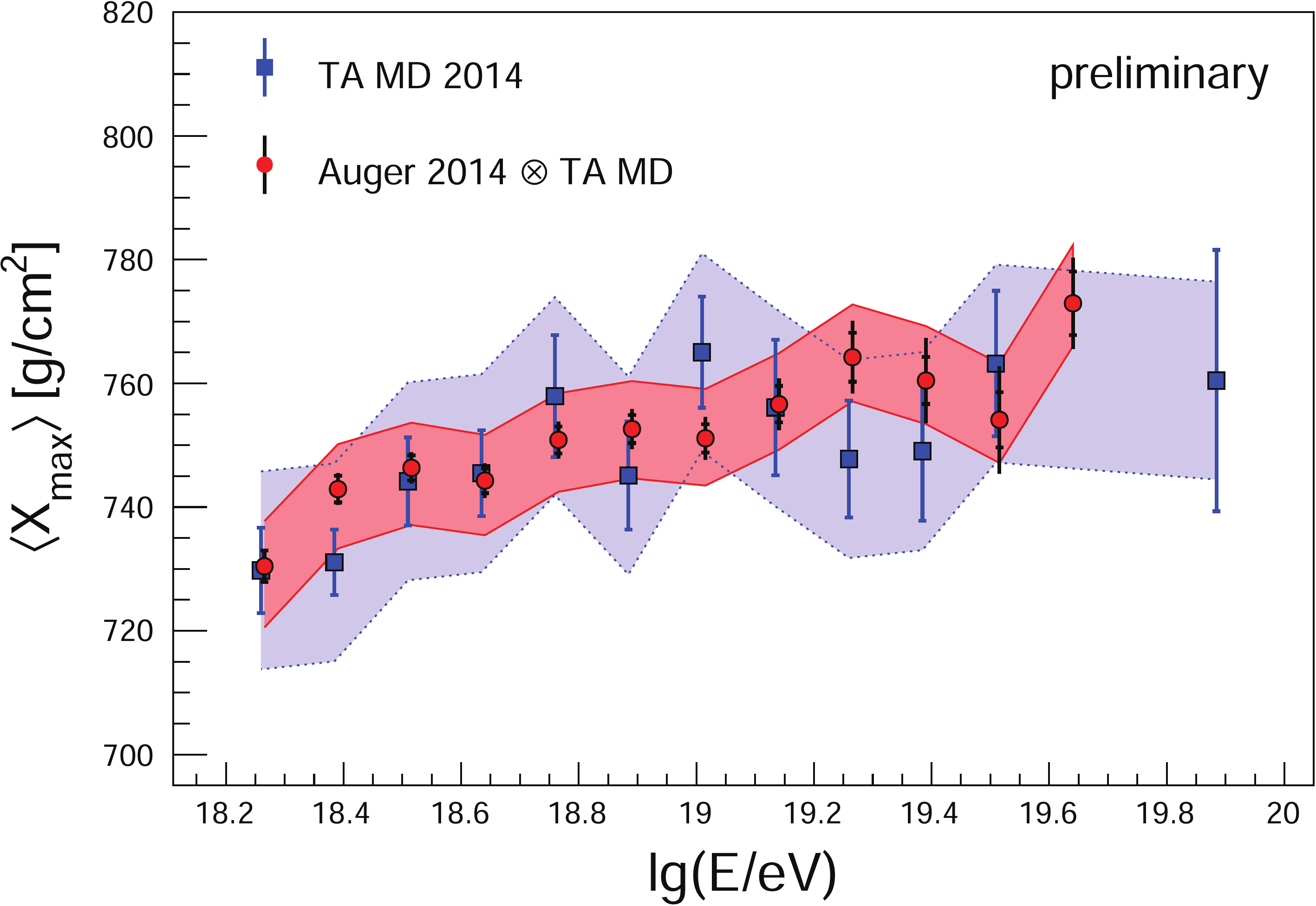}
  \caption{Comparison of the Auger~\cite{longXmax2014} and TA Middle
    Drum~\cite{TA_MD2014} measurements of \meanXmax{}. The plot is taken
    from~\cite{WGmass_UHECR2014}.}
  \label{fig:mdcomp}
\end{figure}

In Fig.~\ref{fig:data_moments_inta} the comparison of
AugerMix~\InTA for \eposlhc (representing the Auger data transferred
into the TA detector) to the TA BR/LR measurements is shown. One can
see that \meanXmax{} of the two experiments agree within the
statistical and systematic errors with mostly shallower \meanXmax{}
values (heavier mass) for TA compared to Auger and that the agreement
is getting better for $E>10^{18.5}$~eV. In the past
(e.g.~\cite{WGmass_UHECR2014}) the agreement of the Auger data on
\meanXmax{} to the Middle Drum TA measurements was tested, and it was
found that both measurements are compatible with an average difference
of $\langle\Delta\rangle=(2.9\pm2.7\,({\rm stat.})\pm18\,({\rm
  syst.}))$~\gcm and no apparent dependence on energy~(see
Fig.~\ref{fig:mdcomp}). The Middle Drum has different hardware and is
located $4-5$~km farther away from the border of the SD compared to
the BR/LR stations and might have worse efficiency for low energy
($E<10^{18.5}$~eV) showers~\cite{TA_ApJ2018_xmax}.

The energy dependence in the comparison of the Auger and TA data can
be naturally tested using the analysis of the measured \meanXmax{}
elongation rates. The elongation rate of Auger is measured to be $(80
\pm 1)$~\gcm/decade between $10^{17.2}$ and $10^{18.3}$~eV, above
$10^{18.3}$~eV it flattens to $(26 \pm 2)$
\gcm/decade~\cite{auger_icrc2017_xmax} indicating the increase of the
primary mass. As noted previously, the TA elongation rate is folded
with the detector effects and thus, due to the detector acceptance, it
can be flatter than expected from the ideal MC. Given the acceptance
biases, the statistical and systematic uncertainties on \meanXmax{}, it
is premature to make interpretations claiming an evolution in
composition from the TA BR/LR data. If \meanXmax{} is fit for all data
observed, the slope is found to be $53 \pm 3$~\gcm/decade, but with
$\chi^2$/ndf of 24/9 ($p\/$-value${}= 4\times10^{-3}$ or $2.7\sigma$
significance) (see also \cite{watson_uhecr2018}). If the lower energy
of the fit is adjusted to $E>10^{18.4}$~eV, the slope is found to be
$40\pm5$~\gcm/decade with an acceptable $\chi^{2}$/ndf of
$5.4/7$. Model predictions of the elongation rate, folded with the
BR/LR detector effects, for a single species is
$\approx50$~\gcm/decade.

Regarding the comparison of \sigmaXmax{}
(Fig.~\ref{fig:data_moments_inta}) up to $10^{18.7}$~eV, there is a
good agreement between the Auger and TA data. Above this energy, the
statistical fluctuations in the TA data become quite large making it
more difficult to draw any firm conclusions on the compatibility of
the Auger and TA measurements.

\section{Discussion}

In this paper the \cwg working group on the mass composition has
presented a method for comparing \xmax data of the two experiments
and the preliminary qualitative checks of the compatibility of
\meanXmax{} and \sigmaXmax{} for energies \energyr{18.2}{19.0} using
the recently published data from the TA BR/LR stations. While we could
not identify any discrepancies beyond the statistical and systematic
errors, more definite conclusions can be obtained after the shower
simulations with \eposlhc for the TA BR/LR will be available. Other
factors which will need to be taken into account are the difference
between the Auger and TA energy scales~\cite{spectrumwg_uhecr2018} and
systematic errors on the fractions of nuclei in the AugerMix MC
templates used by Auger to fit the measured \xmax
distributions. The eventual quantitative comparison will include the
tests of the compatibility of the first two \xmax moments, of the
\xmax distributions and of the \meanXmax{} elongation rates for the
whole energy range $E>10^{18.2}$~eV.

%
%
%
%
%

\end{document}